\documentclass[fleqn,usenatbib,useAMS]{mnras} 

\usepackage{mathptmx} 

\usepackage[T1]{fontenc}
\usepackage{ae,aecompl}

\usepackage{graphicx}   
\usepackage{amsmath}    
\usepackage{balance}    

\newcommand{\Mstar}{\ensuremath{M_\star}}
\newcommand{\Msun}{\ensuremath{\mathrm{M_\odot}}}

\newcommand{\re}{\ensuremath{r_\mathrm{e}}}
\newcommand{\kpc}{\ensuremath{\mathrm{kpc}}}
\newcommand{\Mpc}{\ensuremath{\mathrm{Mpc}}}
\newcommand{\Gpc}{\ensuremath{\mathrm{Gpc}}}

\newcommand{\kms}{\ensuremath{\mathrm{km \  s^{-1}}}}

\title[Relic galaxies prefer dense environments]
      {Massive relic galaxies prefer dense environments}

\author[L. Peralta de Arriba et al.]{Luis
Peralta de Arriba,$^{1,2,3}$\thanks{E-mail: peralta@ing.iac.es (LPdA);
                                            vicent.quilis@uv.es (VQ)}
Vicent Quilis,$^{4,5}$\footnotemark[1]
Ignacio Trujillo,$^{2,3}$
Mar\'{\i}a Cebri\'an$^{2,3}$
\newauthor
and Marc Balcells$^{1,2,3}$
\\
$^{1}$Isaac Newton Group of Telescopes,
      E-38700 Santa Cruz de La Palma, La Palma, Spain\\
$^{2}$Universidad de La Laguna, Departamento de Astrof\'{\i}sica,
      E-38206 La Laguna, Tenerife, Spain\\
$^{3}$Instituto de Astrof\'{\i}sica de Canarias (IAC),
      E-38205 La Laguna, Tenerife, Spain\\
$^{4}$Departament d'Astronomia i Astrof\'{\i}sica, Universitat de Val\`encia,
      E-46100 Burjassot, Val\`encia, Spain\\
$^{5}$Observatori Astron\`omic, Universitat de Val\`encia,
      E-46980 Paterna, Val\`encia, Spain
}

\renewcommand{\today}{24 May 2016} 
\volume{461} 

\date{Accepted 2016 May 20. Received 2016 May 19;
in original form 2016 March 17} 

\pubyear{2016} 

\begin{document}
\label{firstpage}
\pagerange{\pageref{firstpage}--\pageref{lastpage}}
\maketitle

\begin{abstract}
We study the preferred environments of $z \sim 0$ massive relic galaxies ($\Mstar \ga 10^{10}~\Msun$ galaxies with little or no growth from star formation or mergers since $z \sim 2$). Significantly, we carry out our analysis on both a large cosmological simulation and an observed galaxy catalogue.

Working on the Millennium I-\emph{WMAP}7 simulation we show that the fraction of today massive objects which have grown less than 10 per cent in mass since $z \sim 2$ is $\sim$0.04 per cent for the whole massive galaxy population with $\Mstar > 10^{10}~\Msun$. This fraction rises to $\sim$0.18 per cent in galaxy clusters, confirming that clusters help massive galaxies remain unaltered. Simulations also show that massive relic galaxies tend to be closer to cluster centres than other massive galaxies.

Using the New York University Value-Added Galaxy Catalogue, and defining relics as $\Mstar \ga 10^{10}~\Msun$ early-type galaxies with colours compatible with single-stellar population ages older than 10 Gyr, and which occupy the bottom 5-percentile in the stellar mass--size distribution, we find $1.11 \pm 0.05$ per cent of relics among massive galaxies. This fraction rises to $2.4 \pm 0.4$ per cent in high-density environments.

Our findings point in the same direction as the works by Poggianti et al. and Stringer et al. Our results may reflect the fact that the cores of the clusters are created very early on, hence the centres host the first cluster members. Near the centres, high-velocity dispersions and harassment help cluster core members avoid the growth of an accreted stellar envelope via mergers, while a hot intracluster medium prevents cold gas from reaching the galaxies, inhibiting star formation.
\end{abstract}




\section{Introduction} \label{sec:intro}

Using cosmological simulations of galaxy formation, \citet{2013ApJ...773L...8Q} show that a small fraction ($\sim$1 per cent) of $\Mstar \ga 10^{11}~\Msun$ galaxies already formed at $z > 2$ are expected to have grown in mass by less than 10 per cent since that epoch, i.e. they have remained essentially unaltered by mergers, accretion and star formation since $z \sim 2$. This number implies that, in the present-day Universe, $\sim$0.2 per cent of massive galaxies should be relic galaxies, with a number density of $\sim$$10^{-6}~\Mpc^{-3}$. The observational search for $z \sim 0$ massive relic galaxies has been inconclusive to date. \citet{2009ApJ...692L.118T} and \citet{2010ApJ...720..723T}, using the Sloan Digital Sky Survey (SDSS), report only an upper limit of $<10^{-7}~\Mpc^{-3}$ for the number density of massive relic galaxies up to $z \approx 0.1$. In contrast, \citet{2010ApJ...712..226V}, looking in clusters and including lower-mass galaxies, report a large number of massive relic candidates. The latter results raise the question of whether high-density environments (where the SDSS spectroscopic incompleteness is expected to be severe) are particularly favourable for the conservation of massive relic galaxies. It is worth noting that our firmest candidate to be a genuine massive relic galaxy, NGC~1277 \citep{2014ApJ...780L..20T}, inhabits the central region of the Perseus galaxy cluster. This massive cluster \citep[with a velocity dispersion of $1282^{+95}_{-78}~\kms$;][]{1980A&A....82..322D} is known for being the brightest cluster in X-ray emission of the sky \citep{1992MNRAS.258..177E}.

From the theoretical point of view, the location of massive relic galaxies as a function of the environment was explored by \citet{2015MNRAS.449.2396S}. These authors used the characteristics of the dark matter halo of NGC~1277 as a reference for studying, in the BOLSHOI simulation, where galaxies like NGC~1277 are located. They found that dark matter haloes like that of NGC~1277 are substructures of other more massive dark matter haloes. For dark matter haloes corresponding to the richest galaxy clusters, the fraction of sub-haloes similar to the dark matter halo of NGC~1277 can be as high as 30 per cent.

From the observational side, the location of old (with luminosity ages >9 Gyr) superdense galaxies as a function of the environment was explored by \citet{2013ApJ...762...77P}. Using PM2GC \citep{2011MNRAS.416..727C} and WINGS \citep{2010ApJ...712..226V} samples, these authors found that nowadays a significant fraction (at least 17 per cent) of these objects should be found in clusters. They also reported that this result was in agreement with the theoretical expectations derived from their own analysis of the Millenium Simulation. For instance, they found 36 per cent of the galaxies with $\Mstar > 5 \times 10^{10} \  \Msun$ and already passive at $z=1.6$ are members of clusters.

Understanding whether there is any preferential location for massive relic galaxies is key to have a comprehensive picture of the mechanisms responsible of the growth of the massive galaxy population with cosmic time \citep[e.g.][]{2009ApJ...699L.178N,2014MNRAS.444..906F,2015MNRAS.453..704P}. In particular, understanding whether massive relic galaxies inhabit a particular type of environment is key to identifying the physical processes that prevented the size evolution of these objects.

In this paper we provide a new determination of which environment is most favourable for massive relic galaxies. We improve over previous works by conducting a simultaneous analysis of the predictions from numerical simulations and observational data, using the same methodology for characterizing the environment around massive galaxies. In addition, we work with baryonic galaxies instead of just dark matter haloes. Given the low number density of massive relic galaxies in the present-day Universe, we use one of the currently largest $N$-body cosmological simulation: the Millennium I-\emph{WMAP}7 \citep[MI7;][]{2013MNRAS.428.1351G}, and we extend the mass range of the search to lower masses ($\Mstar \ga 10^{10}~\Msun$): there is no reason to believe relics only exist above $\Mstar \ga 10^{11}~\Msun$, and the change improves the statistics that can be extracted from the simulations and the observational data.

The paper is structured as follows. In Section~\ref{sec:sim-cat} we describe our numerical galaxy catalogue, we state the definition of which galaxies will be considered as relics in the simulations (Section~\ref{subsec:sim-relics}), and how we have identified galaxy clusters into them (Section~\ref{subsec:sim-clusters}). In Section~\ref{sec:obs-cat} we introduce our observational catalogue, and define which galaxies will be considered relics within this catalogue (Section~\ref{subsec:obs-relics}). In Section~\ref{sec:env-def} we detail the procedure that we use to characterize the environment of the galaxies in both frameworks (simulations and observations). Our results are shown in Section~\ref{sec:results}, splitting them according they come from simulations (Section~\ref{subsec:sim-results}) or observations (Section~\ref{subsec:obs-results}). In Section~\ref{sec:discussion} we discuss our results, while in Section~\ref{sec:conclusions} we summarize our conclusions.

\section{Numerical galaxy catalogue} \label{sec:sim-cat}

The catalogue of simulated galaxies is based on the public release of a very large $N$-body simulation \citep[MI7;][]{2013MNRAS.428.1351G}. MI7 is a version of the original Millenium I \citep{2005Natur.435..629S} simulation run using the seven-year \emph{WMAP} data \citep{2011ApJS..192...18K}. The cosmological parameters for MI7 simulation are: $\Omega_\mathrm{m}$=0.272, $\Omega_\mathrm{b}$=0.045, $\Omega_\Lambda$=0.728, $n$=0.961, $\sigma_8$=0.807, $H_0$~=~70.4~$\kms$~$\Mpc^{-1}$. The simulation uses $2160^3$ particles in a computational box that has sides of 710~\Mpc. The particle mass is $1.32 \times 10^9~\Msun$.

A combination of two halo finders, a friends-of-friends (FoF) algorithm by \citet{1985ApJ...292..371D} and \textsc{subfind} \citep{2001MNRAS.328..726S}, is used in order to analyse the simulations and to build up the dark matter merger trees. The dark matter haloes found in the $N$-body simulations are transformed into galaxies according to the semi-analytical model of \citet{2013MNRAS.428.1351G}, available in the Millennium data base webpage \citep{2006astro.ph..8019L}.

Guo's model \citep{2013MNRAS.428.1351G} implements several new features with respect to previous ones \citep[i.e.][]{2007MNRAS.375....2D}: the separate evolution of sizes and orientations for gaseous and stellar discs, the size evolution of spheroids, tidal and ram-pressure stripping of satellite galaxies, and the disruption of galaxies to produce intra-cluster light. The effects of AGN feedback are also included. The stellar masses of the semi-analytical galaxies are estimated assuming a \citet{2003PASP..115..763C} initial-mass function (IMF), also used in the observational galaxy catalogue described later in Section~\ref{sec:obs-cat}.

We generate our parent galaxy catalogue using the Millennium data base webpage by selecting all galaxies with $z \sim 0$ stellar masses between $10^{10}$ and $10^{13}~\Msun$.

\subsection{Definition of a relic galaxy in the simulation} \label{subsec:sim-relics}

We define a $z \sim 0$ relic in the simulation as a galaxy that has barely increased its stellar mass since $z \sim 2$. In order to identify the possible candidates to relic galaxies in our numerical catalogue, we use the merger tree structures to trace backwards in time the massive galaxies identified at $z \sim 0$, together with two conditions.
\begin{enumerate}
\item The galaxy must be already formed at $z \sim 2$, namely, it must be identified by the halo finder algorithm as a group of gravitationally bound particles. In our particular implementation, and in order to ensure that the galaxies are numerically well resolved, only galaxies with stellar masses larger that $10^9 \ \Msun$ are considered.
\item The stellar mass of the $z \sim 2$ precursor is higher than the 90 per cent of the $z \sim 0$ mass.
\end{enumerate}

\subsection{Identification of galaxy clusters in the simulation} \label{subsec:sim-clusters}

The next step in our procedure uses a simplified version of the idea of the spherical over-density halo finders \citep[see e.g.][]{2009MNRAS.399..410P} to identify the possible candidates to be a virialized galaxy cluster in the sample. The basic concept of this technique is to find spherical regions with an average density above a certain threshold, which can be fixed according to the spherical top-hat collapse model. Therefore, we can define the virial mass of a halo ($M_\mathrm{vir}$) as the mass enclosed in a spherical region of radius ($r_\mathrm{vir}$) having an average density $\Delta_\mathrm{c}$ times the critical density $\rho_\mathrm{c} (z) = 3H(z)^2 / 8 \pi G$:
\begin{equation} \label{eq:mvir}
M_\mathrm{vir} (<r_\mathrm{vir}) =
\frac{4}{3} \pi r_\mathrm{vir}^3 \Delta_\mathrm{c} \rho_\mathrm{c}.
\end{equation}

The over-density $\Delta_\mathrm{c}$ depends on the adopted cosmological model, and can be approximated by the following expression \citep{1998ApJ...495...80B}:
\begin{equation} \label{eq:deltac}
\Delta_\mathrm{c} = 18 \pi^2 + 82 x - 39 x^2,
\end{equation}
where $x = \Omega(z) - 1$ and $\Omega(z) = [\Omega_\mathrm{m} (1 + z)^3]/[\Omega_\mathrm{m} (1 + z)^3+ \Omega_\Lambda]$. Depending on the different cosmologies, typical values of $\Delta_\mathrm{c}$ can vary between 100 and 500. For the particular choice of cosmological parameters of this paper, the value of $\Delta_\mathrm{c}$ obtained using equation~(\ref{eq:deltac}) is $\Delta_c \simeq 350$.

In our particular implementation of a spherical over-density finder, and given the fact that we do not consider total masses but stellar masses, we use $\Delta_\mathrm{c}$ to define a virialized region that has a stellar average density 350 times the average stellar density in the whole computational domain. For these computations, we consider the stellar masses from massive galaxies (i.e. those above $10^{10} \ \Msun$). With these conditions, we look for the galaxy whose spherical region has the highest average density. If this quantity divided by the average stellar density in the whole simulation is above the threshold $\Delta_\mathrm{c} \simeq 350$, we define this region as a galaxy cluster. The procedure is repeated looking for the next galaxy whose associate sphere has a environmental density fulfilling the density threshold condition and non-overlapping with previously identified clusters. The procedure goes on until no candidates to galaxy cluster satisfying the conditions are left. Once all the galaxy clusters are identified, we characterize them by computing their velocity dispersion ($\sigma$) using the relative velocities with respect to the cluster centre of mass for all the galaxies in the cluster.

\section{Observational galaxy catalogue} \label{sec:obs-cat}

Our sample of real galaxies is obtained from the New York University Value-Added Galaxy Catalogue \citep[hereinafter NYU-VAGC;][]{2005AJ....129.2562B}, which is based in the data release 7 of the Sloan Digital Sky Survey \citep[hereafter SDSS-DR7;][]{2009ApJS..182..543A}. This catalogue contains about $2.5 \times 10^6$ objects with spectroscopic redshifts. In addition to photometric and spectroscopic information extracted from SDSS-DR7, the NYU-VAGC provides structural parameters, $K$-corrections and stellar mass estimates \citep[derived using a \citealt{2003PASP..115..763C} IMF;][]{2007AJ....133..734B}. Tabulated quantities that depend on the adopted cosmology were converted to the cosmology adopted in MI7 and \emph{WMAP} (Section~\ref{sec:sim-cat}).

In order to avoid edge effects when exploring the environment of our real galaxies, we only consider the galaxies which live in the largest continuous volume of SDSS-DR7. In particular, we stay within the following region, which was originally proposed by \citet[][cf. their section 2.2 and their fig.~2]{2012ApJ...744...82V}:
\begin{enumerate}
\item Southern edge: $\delta > 0$.
\item Western edge: $\delta > 2.555556 (-\alpha + 131\degr)$.
\item Eastern edge: $\delta > 1.70909 (\alpha - 235\degr)$.
\item Northern edge: $\delta < \arcsin\left[\frac{0.93232 \sin(\alpha - 95.9\degr)}{\sqrt{1 - [0.93232 \cos(\alpha - 95.9\degr)]^2}}\right]$.
\end{enumerate}

Also, to get a mass-complete sample and guarantee that our definition of environment is homogeneous over the redshift interval of exploration, we constrain our sample to galaxies with redshifts between 0.005 and 0.060. According to equation~(1) of \citet[][cf. also their figs~3 and 4]{2014MNRAS.444..682C}, this redshift interval ensures that our sample will be complete for galaxies with stellar masses above $10^{10}~\Msun$.

Selection biases of the SDSS spectroscopic catalogue affecting nearby, high-surface brightness galaxies have been carefully studied by \citet{2010ApJ...720..723T}. Specifically, they analysed the effects of the star/galaxy separation criterion, the `saturation' selection limit and the `cross-talk' selection limit. For the range of redshift of our sample, the most important effect is the `cross-talk' selection limit. To first order, these selection biases are independent of environmental density, and consequently it is not expected that they affect our observational results. Nevertheless, and in order to show the robustness of these results, we will discuss about this issue in Section~\ref{sec:discussion}.

With the above restrictions, our observational catalogue is a mass-complete sample of 41716 galaxies with $\Mstar > 10^{10}~\Msun$. From this sample, we classify 27034 objects (65 per cent) as early-type galaxies on the basis of their S\'ersic indices (i.e. $n > 2.5$), while 14682 objects (35 per cent) are considered late-types (i.e. $n < 2.5$).

\subsection{Definition of a relic galaxy in the observations} \label{subsec:obs-relics}

Identifying a relic galaxy in the observations is less straightforward than in simulations, as we have poor information on the merger and the star-formation histories of real galaxies. Using the criterion that a relic galaxy should not have undergone star formation since $z \sim 2$ and should not have evolved structurally since those redshifts, we require three conditions to classify a galaxy as a candidate relic:
\begin{enumerate}
\item Its stellar population is old: equivalent single-stellar-population (SSP) age above 10 Gyr.
\item Its structure is early-type: with S\'ersic index $n > 2.5$.
\item It is compact: its size must be close to stellar mass--size relationship of early types at $z \sim 2$.
\end{enumerate}

To identify old galaxies, we select the reddest galaxies in the rest-frame colour--colour diagram $g-r$ versus $r-i$. We base our colour cut using a SSP model with age 10 Gyr and metallicity $-0.25$. The values adopted for this SSP were taken from the photometric predictions based on MIUSCAT spectral energy distributions \citep{2012MNRAS.424..172R,2010MNRAS.404.1639V,2012MNRAS.424..157V}. Hence, we require $g-r > 0.763$ and $r-i > 0.363$. Additionally, and in order to minimize the contamination of dusty galaxies in our sample of relic galaxies, we reject the most asymmetric objects on the sky, i.e. those with a ratio between minor and major axes lower than 0.3, because they are the most probable candidates to be galaxies with dusty discs. We will expand about the contamination of dusty galaxies and the effectiveness of our rejection criteria on the discussion exposed in Section~\ref{sec:discussion}.

The structural requirements derive from the properties of massive galaxies at high redshift. Many works have shown that massive galaxies have grown with cosmic time \citep[e.g.][]{2005ApJ...626..680D,2007MNRAS.382..109T,2008ApJ...687L..61B}, being the size evolution stronger for early-type galaxies. Furthermore, stellar mass--size scaling relationship seems to have a constant slope with cosmic time. This result has been reported by \citet{2014ApJ...788...28V} up to $z \sim 3$, finding that the major-axis radius $R_\mathrm{e,maj}$ is proportional to $\Mstar^\beta$ with $\beta \sim 0.75 \pm 0.05$ for early-type galaxies. In the nearby Universe the exponent is just slightly shallower. \citet{2003MNRAS.343..978S} found that nearby early-type galaxies (i.e. with S\'ersic index $n > 2.5$) are well described by a scaling relationship between their effective radii and stellar masses of the type $\re \propto \Mstar^\alpha$ with $\alpha = 0.56$.

Taking advantage of the stronger global size evolution of early-type galaxies with cosmic time, we select our candidates for being relic galaxies as those objects with old stellar populations and S\'ersic index $n$ above 2.5 located in the lower region of the stellar mass--size plane. In particular, we define this lower region to contain 5 per cent of the whole sample of galaxies with the smaller sizes at a each stellar mass. This means we consider as compact galaxies those with sizes below
\begin{equation} \label{eq:re}
\re = \lambda \left(\frac{\Mstar}{10^{10} \ \Msun}\right)^{0.56},
\end{equation}
being $\lambda = 0.749~\kpc$ and $\re$ the effective radius in the $r$ band.

\begin{figure}
    \includegraphics{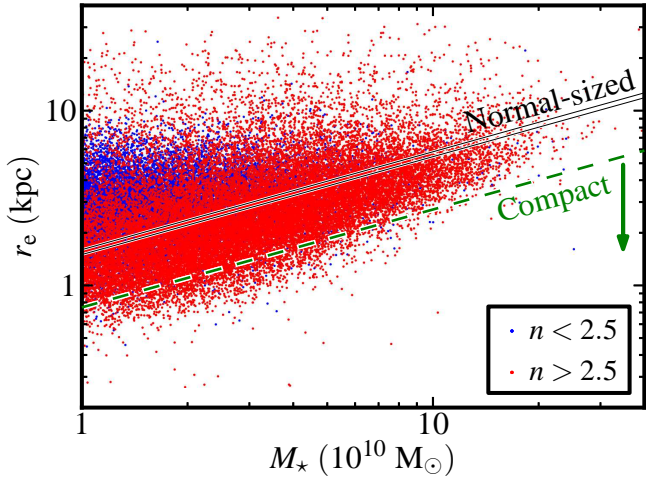}
    \caption{From the observational catalogue, distribution over the stellar mass--size plane of the galaxies. The green dashed line indicates the upper limit of the compact definition given in Section~\ref{subsec:obs-relics}, while the two solid black lines show the region populated by the galaxies labelled as normal-sized in the same section. In addition, we have represented the galaxies with different colours depending on their S\'ersic indices as indicated in the legend.}
    \label{fig:obs-mass-size}
\end{figure}

\begin{figure}
    \includegraphics{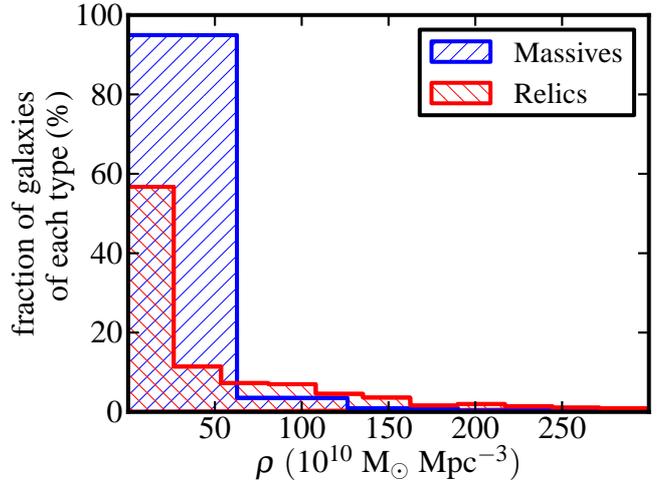}
    \caption{From the numerical catalogue, distribution of the environmental density of massive (blue) and relic (red) galaxies.}
    \label{fig:sim-env}
\end{figure}

We define a comparison sample of normal-sized galaxies by taking all galaxies between the 47.5 and 52.5 percentiles of the distribution $\re / \Mstar^{0.56}$ of the whole sample (i.e. without any restriction on S\'ersic-index). They are the galaxies which are between two lines in the stellar mass--size plane of the form of equation~(\ref{eq:re}) with values of $\lambda$ equal to $\lambda_1 =1.502~\kpc$ and $\lambda_2 = 1.602~\kpc$. In Fig.~\ref{fig:obs-mass-size} we show how compact and normal-sized definitions are reflected in the stellar mass--size plane.

\section{Characterizing the density of the environment} \label{sec:env-def}

The same procedure is used to define environmental density in the simulations and in our real-galaxy catalogue. Following a similar prescription to that described in \citet{2014MNRAS.444..682C}, we define the environmental density around each galaxy as follows: for each galaxy we identify all its neighbours with masses above $10^{10}~\Msun$ within a sphere of radius $R = 0.5~\Mpc$, and define the environmental density as follows:
\begin{equation} \label{eq:rho}
\rho_i = \frac{1}{\frac{4}{3} \pi R^3} \sum_{k=1}^{N_i} M_{i,k},
\end{equation}
where $M_{i,k}$ is the stellar mass of the $k$th neighbour located at less than $R$ of the $i$th galaxy of the sample (galaxy $i$ has $N_i$ neighbours). The selected radius of $R = 0.5~\Mpc$ is a compromise between having a local measurement of the density and at the same time having a volume large enough to guarantee that there is a significant number of galaxies above the stellar mass completeness limit so we can compute a sensible density around our targeted galaxy.

Although the same formal definition is used for both frameworks, density values from simulations and observations cannot be directly compared to each other. In the case of the real-galaxy catalogue, because the line-of-sight coordinate is inferred from the redshift, high-density virialized regions will be stretched along the line of sight (fingers-of-god effect) hence equation~(\ref{eq:rho}) provides a lower-limit to the actual physical density. However, it is worth noting that the computed densities for real galaxies have been proved to be a useful proxy to trace overdensities \citep[e.g.][]{2005ApJ...634..833C,2012MNRAS.419.2133H,2012MNRAS.419.2670M,2014MNRAS.444..682C}. To emphasize this difference, we name the inferred quantity $\tilde{\rho}$ when it has been computed from the real-galaxy catalogue.

\section{Results} \label{sec:results}

\begin{figure}
    \includegraphics{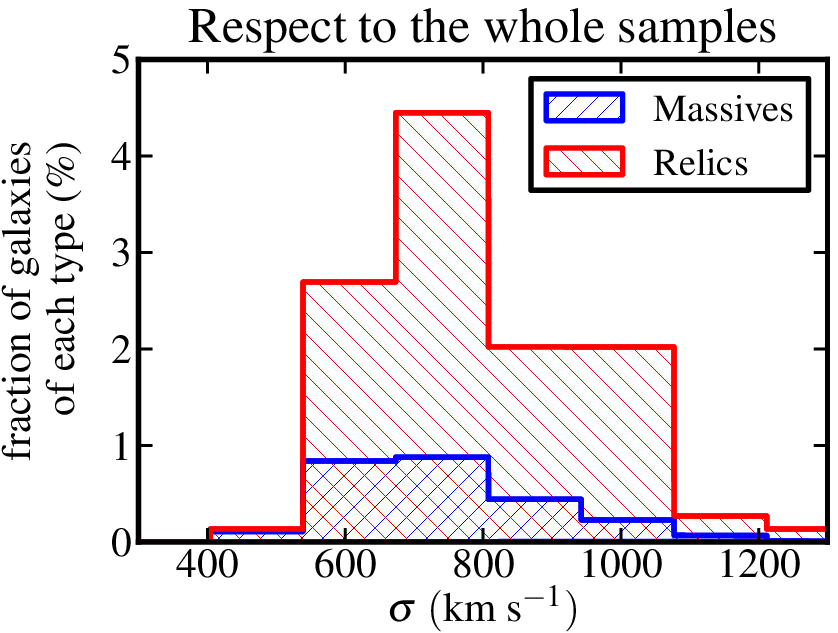}
    \includegraphics{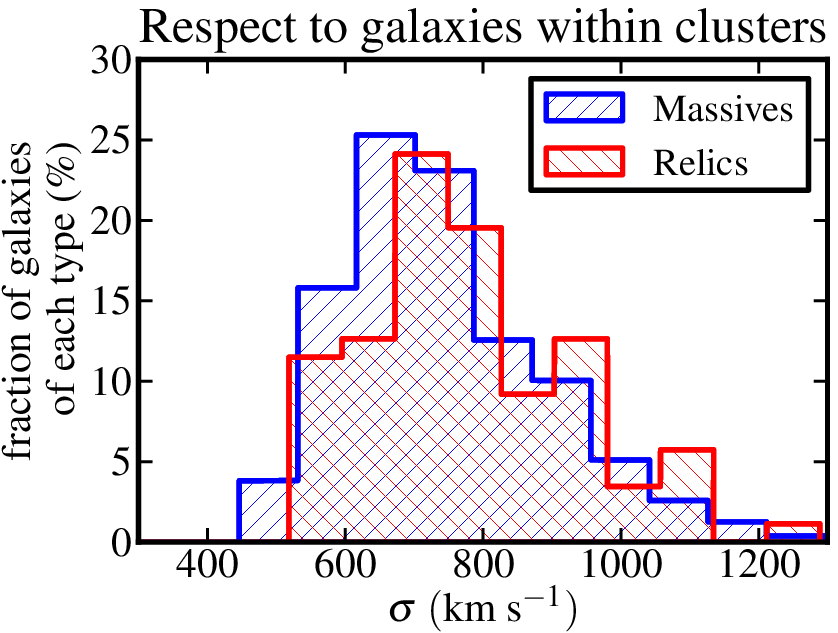}
    \caption{From the numerical catalogue, fraction of galaxies as a function of the velocity dispersion of the host galaxy cluster. Upper panel shows the fraction of each galaxy type with respect to the whole sample of massive galaxies (blue) and relics (red). Lower panel shows fractions referred to the amount of galaxies of each type which live within clusters.}
    \label{fig:sim-clusters-sigma}
\end{figure}

\begin{figure}
    \includegraphics{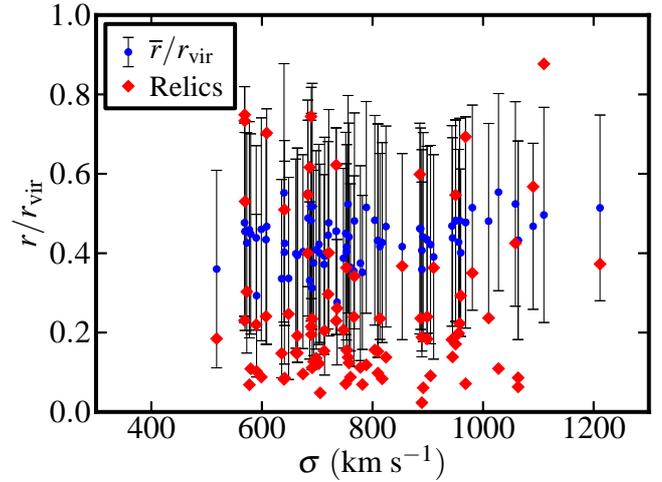}
    \caption{From the numerical catalogue, normalized radial position of relic galaxies in clusters (red diamonds) as a function of the velocity dispersion of the clusters, compared with the averaged radial positions $\overline{r}$ of all galaxies in each cluster (blue circles). Error bars represent one standard deviation for the radial distribution of the galaxies in a given cluster.}
    \label{fig:sim-clusters-radius}
\end{figure}

\subsection{The location of relic galaxies according to the simulations} \label{subsec:sim-results}

We use the catalogue of galaxies with stellar masses larger that $10^{10}~\Msun$ extracted from the Millennium simulation. In this sample, we identify 1850648 massive galaxies ($\Mstar > 10^{10}~\Msun$) at $z \sim 0$. Among the subsample of massive galaxies, 742 galaxies can be labelled as relics according to the criteria described in Section~\ref{subsec:sim-relics}. This number of relic galaxies corresponds to an extremely tiny fraction of today galaxies with $\Mstar > 10^{10}~\Msun$: 0.04 per cent. This value is lower than the measurement reported by \citet{2013ApJ...773L...8Q}, where they found that the number of present-day massive relic galaxies is 0.06 per cent when they used our same catalogue \citep[from][]{2013MNRAS.428.1351G}. We have checked that the above difference is only due to the higher mass cutoff ($\Mstar > 8 \times 10^{10}~\Msun$) of the \citet{2013ApJ...773L...8Q} sample.

We now direct our attention to the number of relic galaxies in clusters in the simulations. By using the method described in Section~\ref{subsec:sim-clusters}, we identify 672 rich galaxy clusters. Within these clusters, we find 47646 massive galaxies and 87 relics (i.e. 0.18 per cent of them). Consequently, despite that the simulation has 8.5 times fewer relic galaxies in clusters than in the field, it follows that the probability for a massive galaxy to be a relic is 4.6 higher in clusters than in a global search of the population of massive galaxies.

\subsubsection{The distribution of relic galaxies as a function of the environmental density}

To make the analysis more quantitative, we explore the distribution of massive galaxies (relic and not) as a function of the environmental density as defined in equation~(\ref{eq:rho}). The distribution of $\rho$ for the galaxies in the simulation is shown in Fig.~\ref{fig:sim-env}. We see that the distribution of relics (red line) shows a more extended tail towards higher values than that of non-relic massive galaxies (blue line). This result points in the same direction as the above findings, where we report that the fraction of relics is higher in clusters.

\subsubsection{Relics in clusters: velocity-dispersion distribution}

In this section we explore how the massive (relic or not) galaxies are distributed according to the velocity dispersion of the clusters (i.e. a proxy of the galaxy cluster global mass). The fraction of massive galaxies in clusters with respect to the total population of massive galaxies in our simulation is 2.6 per cent. Their distribution within the galaxy clusters is shown in upper panel of Fig.~\ref{fig:sim-clusters-sigma}. Similarly, the fraction of relic galaxies in clusters with respect to the total population of relic galaxies in our simulation is 11.7 per cent and their distribution is shown in the same panel. In the upper panel of Fig.~\ref{fig:sim-clusters-sigma} the higher area from the histogram of relic galaxies reflects the higher probability for a massive galaxy to be a relic in rich clusters, while the ratio between the two histograms would give an analogue probability referred to each velocity-dispersion bin.

If we focus now our attention to the distributions considering the fractions with respect to the amount of galaxies of each type within the clusters, we can appreciate better whether there is any difference in the distribution of relic galaxies as a function of the velocity dispersion of the clusters. Bottom panel of Fig.~\ref{fig:sim-clusters-sigma} shows the result of this exercise. There is a small shift of the relic galaxies to higher velocity-dispersion values compared to the distribution of the other massive galaxies, i.e. the median in the distribution of the relics is 33 \kms\ higher. In order to guess if this difference is statistically significant, we performed a Kolmogorov--Smirnov test using both samples. Our result was a $p$-value of 0.055, which means that both distributions are marginally different.

\subsubsection{Relics in clusters: radial distribution}

Focusing on the subsample of galaxies in clusters, we enquire whether relic galaxies show any particular preference in their location within the cluster. Fig.~\ref{fig:sim-clusters-radius} presents the radial location of the relics (red diamonds) in clusters against the cluster velocity dispersion. For the same clusters where the relics are located, we derive the mean $\overline{r}$ and the standard deviation of the radial distribution of all galaxies in the clusters. These are shown with blue circles with error bars. Comparing the radial location of the relics with the average position of the galaxies, we see that most of the relics are lower than the second ones, i.e. relic galaxies used to be a factor of 2 closer to the cluster centres.

\subsection{The location of relic galaxies in the observed catalogue} \label{subsec:obs-results}

\begin{figure}
    \includegraphics{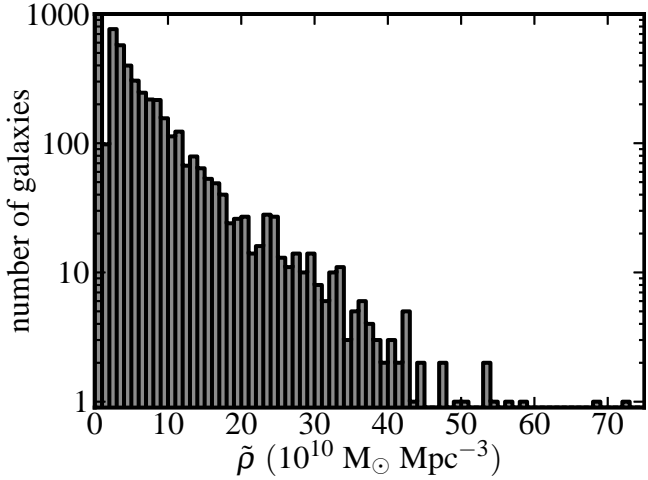}
    \caption{From the observational catalogue, distribution of the environmental density of the massive galaxies. The bin size is $10^{10} \ \Msun \ \Mpc^{-3}$. The first bin has been cut in this figure, but it contains 37847 objects and corresponds to the galaxies without neighbours in our sample.}
    \label{fig:obs-env}
\end{figure}

\begin{figure}
    \includegraphics{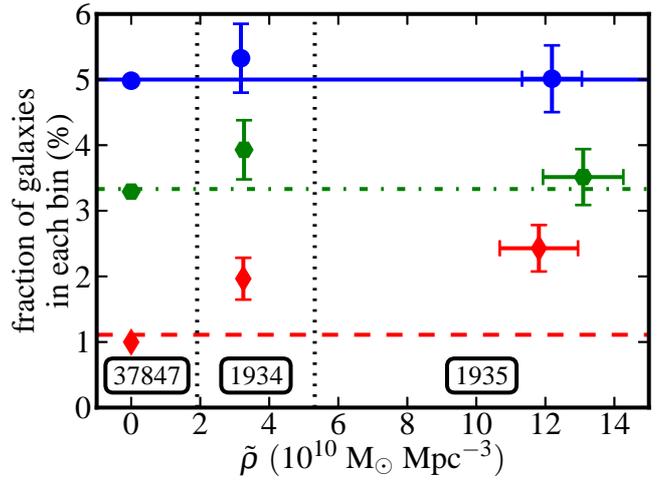}
    \caption{From the observational catalogue, fraction of normal-sized galaxies (blue circles), normal-sized early-type galaxies (green hexagons) and relics (red diamonds) in each environmental-density bin for the subsample. Horizontal blue solid, green dash--dotted and red dashed solid lines show the global fractions of galaxies (i.e. those computed neglecting any $\tilde{\rho}$ binning) which are normal-sized galaxies, normal-sized early-type galaxies and relics, respectively. Vertical black dotted lines show the limits used for binning in the $\tilde{\rho}$ axis, while the numbers within boxes indicate the number of galaxies of the whole sample which belong to each $\tilde{\rho}$ bin. The error bars with similar sizes to the symbol sizes were omitted for clarity.}
    \label{fig:obs-fraction-env}
\end{figure}

We now shift our attention to the observational catalogue, and enquire on the abundance and environmental density distribution of relic galaxies. Out of the 41716 massive galaxies in the catalogue, we find 463 galaxies classified as relics according to the criteria described in Section~\ref{subsec:obs-relics}. The global relic fraction is therefore $1.11 \pm 0.05$ per cent. With respect to the normal-sized galaxies that we will use as reference, we find 2086 objects (5.0 per cent of the sample by definition), 1390 of them being early-type galaxies (3.3 per cent). 

In Fig.~\ref{fig:obs-env} we show the distribution of environmental density for all massive galaxies from the observational catalogue defined in Section~\ref{sec:obs-cat}. Densities range from 0 to $72 \times 10^{10} \ \Msun \ \Mpc^{-3}$. We note that this range is smaller than that found in simulated relics, which reaches values up to $\rho \la 300 \times 10^{10} \ \Msun \ \Mpc^{-3}$ (cf. Fig.~\ref{fig:sim-env}). The main reason for this difference is the fingers-of-god effect, which was already expected from the reasons exposed when we presented our procedure to estimate the environmental density in Section~\ref{sec:env-def}. In addition, the large difference between the volumes of observational ($0.011~\Gpc^3$) and numerical ($0.358~\Gpc^3$) catalogues contributes to the difference in maximum densities. Also contributing must be the spectroscopic redshift incompleteness of the SDSS: the SDSS spectroscopic catalogue is claimed to be $\sim$99-per-cent complete down to $r\sim17.77~\mathrm{mag}$ \citep{2002AJ....124.1810S}. This will turn out into having not only a smaller number of massive galaxies in the real catalogues but also an artificially smaller environmental density around them, especially in the densest regions.

Fig.~\ref{fig:obs-fraction-env} shows the fraction of normal-sized galaxies (blue circles), normal-sized early-type galaxies (green hexagons) and relics (red diamonds) in three $\tilde{\rho}$ bins. The first bin contains 37847 isolated galaxies of our sample (i.e. those with $\tilde{\rho} = 0$), while the other two bins were chosen to contain a similar number of objects (1934 and 1935 galaxies). The limits between these $\tilde{\rho}$ bins are indicated with vertical dotted lines in the figure. Horizontal lines are used to show the fraction of each type of galaxy without considering any $\tilde{\rho}$ binning (in particular, solid blue/green dash--dotted/red dotted line corresponds to the fraction of normal-sized/normal-sized early-type/relic galaxies). Fig.~\ref{fig:obs-fraction-env} reveals that the fraction of relic galaxies is a monotonic increasing function of the environmental density. In contrast, galaxies with average sizes in the stellar mass--size relationship are equally distributed as a function of the environmental density, even when we focus only on normal-sized early-type galaxies. The increasing trend shown by the relic galaxies in Fig.~\ref{fig:obs-fraction-env} would be probably more accentuated if SDSS were not affected by incompleteness. The reason is that this incompleteness is more important in the centres of the galaxy clusters due to fibre collision problems \citep[e.g.][]{2012ApJ...756..127G}.

The trend shown by Fig.~\ref{fig:obs-fraction-env} obviously depends on the size $R$ of the fixed-aperture sphere considered in the definition of the environmental density. In this work we have considered $R = 0.5 \ \kpc$. We checked that for radii bigger than 2 \Mpc\ the trend disappears, while it is weaker (but still present) for 1 \Mpc. As the size of the sphere for characterizing the environmental density increases our ability to describe the local conditions also decreases. Consequently, we can conclude that the physical conditions for maintaining the galaxies in their initial (relic) configuration are linked with the local (i.e. sub-cluster) scales.

\section{Discussion} \label{sec:discussion}

Both the simulation and the observational data compiled in this work lead to the same conclusion: massive ($\Mstar > 10^{10}~\Msun$) relic galaxies can be found in all kind of environments; however, their relative fraction increases with the environmental density. According to the simulations the difference in the relative fraction of relic galaxies in clusters compared to the field can be as high as 4.6. The results that we find in this work suggest that the local environment plays a major role in determining when a massive galaxy formed at $z \ga 2$ is not going to evolve much since its formation. It is particularly relevant for this discussion that, in the simulations, the location of relic galaxies in clusters peaks towards the central parts. The central regions of clusters are also expected to form and virialize early, indicating that the mechanism necessary for the subsequent growth of massive galaxies was slowed down very early on.

The centres of the clusters, particularly the most massive ones, are characterized for having a large velocity dispersion and also a very hot intracluster medium. These conditions prevent an effective mechanism for mass (and size) growth of the galaxies. If the picture drawn here is correct, the right places to find a significant number of massive relic galaxies are on the centre of the clusters. Observationally, to claim the detection of a genuinely relic galaxy is not easy. One has to guarantee that the stellar populations are old throughout the entire structure of the galaxy and this is observationally very demanding. So far, this study has been only conducted in NGC~1277 \citep[][]{2014ApJ...780L..20T}. Interestingly, this massive and compact galaxy is very closely located towards the centre of the very rich Perseus cluster.

The argument of the previous paragraph focused on how the clusters can help to prevent the galaxy growth and the \emph{rejuvenation} of its stellar population through star formation. Nevertheless, due to hierarchical evolution we could also expect that relics prefer clusters. Since relic galaxies are a subset of galaxies which were formed first (by definition), they have had more time to cluster than galaxies formed later. We have to take into account that even in this scenario, the key to the survival of these relics resides in the large velocity dispersions and the hot intracluster medium.

We would like to point out that the previous interpretation includes our observational findings under the assumption that high-density environments should correspond in their majority to clusters. This is justified as several works \citep[e.g.][]{2012MNRAS.419.2133H,2012MNRAS.419.2670M} have shown that the aperture-based methods are a good proxy for the halo mass and, consequently, for selecting cluster environments.

Comparing our results with previous (observational and numerical) works, we find a nice agreement. As commented in the Introduction (Section~\ref{sec:intro}), \citet{2015MNRAS.449.2396S}, simulating the dark matter component only, also obtain in the BOLSHOI simulation, an increasingly large number of relic galaxies inhabiting the most massive galaxy clusters. From the observational side, \citet{2010ApJ...712..226V} find a large number of massive and compact galaxies in clusters.

The first clues indicating that relic galaxies should be searched in clusters were given by \citet{2013ApJ...762...77P}. In the nearby Universe, they observed that at least 17 per cent of old (with luminosity ages >9 Gyr) superdense galaxies should be found in clusters. Using the Millenium Simulation, they found 36 per cent of today galaxies with $\Mstar > 5 \times 10^{10} \ \Msun$ and already passive at $z=1.6$ are members of clusters. In our work, 12 per cent of the relics from the simulations are part of rich clusters, while in the observations 10 per cent of the relics fall in the highest $\tilde{\rho}$ bin (cf. Fig.~\ref{fig:obs-fraction-env}). Hence, we broadly coincide with \citet{2013ApJ...762...77P} in finding a significant fraction of relics in clusters. The quantitative differences between the fractions inferred in the two studies, and between the fractions inferred from observations and from simulations in each study, are to be expected from differences in the adopted definitions of relic and environmental density.

The reason why \citet{2009ApJ...692L.118T} and \citet{2010ApJ...720..723T} failed to identify relic galaxies in the SDSS appears to be the more stringent mass ($\Mstar > 10^{11}~\Msun$) and compactness constraints used in their search: the galaxy mass function drops steeply at these masses, and catalogues simply have fewer galaxies with $\Mstar > 10^{11}~\Msun$. The known spectroscopic incompleteness of SDSS towards the centre of the clusters also contributes to the negative result reported in the above two papers. In fact, when the stellar mass and size of the galaxies are a little bit relaxed, a good number of candidates for being relic galaxies based on the velocity dispersion of the objects is found \citep{2015A&A...578A.134S}.

It will be worth exploring the environment of relic galaxies at different redshifts. Many recent works have characterized the number density of massive relic galaxies against redshift \citep{2014ApJ...793...39D,2014ApJ...780..134S,2014ApJ...796...92H,2015ApJ...806..158D,2016MNRAS.457.2845T}. The next step in this direction is to probe whether the relative abundance of those intermediate-redshift relic galaxies is higher in denser environments. The findings in the present paper predict that they will be. A very recent work by \citet{2015ApJ...815..104D} seems to support this prediction.

Could the dependency of the relic fraction with environmental density have been inferred from trends of the stellar mass--size relationship with density? We do not think so. The variation with environment and redshift of the stellar mass--size relationship for early-type/quiescent galaxies has been studied \citep[e.g.][]{2013MNRAS.428.1715H,2013ApJ...779...29H,2014MNRAS.444..682C}, and their distributions, characterized by parameters such as mean/median or standard deviation, contain insufficient information about the bottom tail of the distributions for characterizing the relic fraction.

We refer in Section~\ref{sec:obs-cat} to the incompleteness of the SDSS spectroscopic catalogue in nearby massive galaxies studied by \citet{2010ApJ...720..723T}. These authors warn about biases against massive compact galaxies at low redshifts ($z \la 0.05$). Here explore whether the trends outlined in this paper might be the result of such biases. To that effect, we split our sample in two redshift ranges: $z \la 0.05$ where bias effects might be higher, and $z \ga 0.05$ where bias effects are expected to be weaker or non-existent. From the observational volumes of each of the two sub-samples, we estimate that we could be missing approximately 140 galaxies in the low-redshift range due to these biases. The strongest source of bias comes from saturation limits and hence from high surface brightness; it must show very little dependency on environment, and hence there is no reason for expecting that the missed relics show a different environment distribution than the galaxies of our sample. This leads us to conclude that the effects of this bias on the trends of relic fraction with density must be very small. But even if all of the 140 missed galaxies would live in $\tilde{\rho}=0$ environments, we find that the observational trend of a higher fraction of relics in higher-density environments remains. Additionally, we note that incompleteness in the SDSS spectroscopic catalogue is in fact expected to be higher in clusters due to fibre proximity limits. The effect of such biases would be to dilute, rather than strengthen, the trends found in this paper.

Additionally, we would like to comment that we also checked the dependence of our results on the method that we have used to identify clusters in the simulations. In particular, we have also extracted the list of haloes with stellar masses larger than $10^{14} \ \Msun$ using an FoF method in the Millennium simulation. In this case, the number of relic galaxies in the clusters detected by the FoF method is totally comparable to the one found by the method used in this work (described in Section~\ref{subsec:sim-clusters}).

Finally, we would like to mention that the contamination of dusty galaxies in the observational relic sample should be tiny. As it was explained in Section~\ref{subsec:obs-relics}, due to the unavailability of infra-red data to reject the dusty galaxies, we decided to exclude from our sample the most asymmetric objects in the sky in order to avoid including galaxies which potentially could contain edge-on dusty discs. Applying this filter, we rejected 13 galaxies. Assuming a random orientation of the dusty discs in the sky, and taking into account that the ratio between minor and major axes used in our filter is 0.3, we should have rejected around 70 per cent of the dusty discs. This means that within the considered sample there should be around six face-on dusty discs, which is a small fraction of our sample of relic galaxies. It is also worth noting that this method is conservative, and therefore it overestimates the contamination. Furthermore, we would like to mention that the inclusion of the potentially edge-on dusty galaxies does not change the observational results shown in this work.

\section{Conclusions} \label{sec:conclusions}

The main conclusions of this paper can be summarized as follows.
\begin{enumerate}
\item Cosmological simulations predict that the fraction of present-day relic galaxies with $\Mstar > 10^{10}~\Msun$ is only 0.04 per cent. This number is 4.6 higher when the galaxies are located in cluster environments: 0.18 per cent. Moreover, it seems that this fraction could be even larger in the most massive clusters with large velocity dispersions (although the statistical significance of this last result is marginal).
\item The distribution of the relic galaxies in the simulation within the clusters is not homogeneous. In fact, they tend to be located a factor of 2 closer to the central parts of the clusters than the other massive galaxies. These very dense regions slow down the physical mechanisms for the mass growth of the galaxies.
\item Observationally, in agreement with the simulations, we also find that the fraction of relic galaxies increases as their surrounding densities rise.
\end{enumerate}

These conclusions point in the same direction as previous works: the article by \citet{2013ApJ...762...77P}, which reported the preference of old (with luminosity ages >9 Gyr) superdense galaxies to inhabit in clusters; and the article by \citet{2015MNRAS.449.2396S}, which found that dark matter haloes like that of the relic NGC~1277 are substructures of other more massive dark matter haloes.



\section*{Acknowledgements}

The authors are grateful to the referee for his/her useful and constructive review. The authors thank A.~Vazdekis and I.~Mart\'{\i}n-Navarro for their collaboration during the development of this paper. This work has been supported by the Programa Nacional de Astronom\'{\i}a y Astrof\'{\i}sica of the Spanish Ministry of Economy and Competitiveness (under the grants AYA2009-11137, AYA2013-48226-C3-1-P and AYA2013-48226-C3-2-P) and the Generalitat Valenciana (under the grant PROMETEOII/2014/069). The Millennium Simulation data bases used in this paper and the web application providing online access to them were constructed as part of the activities of the German Astrophysical Virtual Observatory. This research made use of \textsc{astropy}, a community-developed core \textsc{python} package for Astronomy \citep{2013A&A...558A..33A}.

\label{lastpage} 
\balance 




\bibliographystyle{mnras}
\footnotesize{\bibliography{peralta_de_arriba_quilis_et_al-2016}} 


\bsp
\end{document}